\documentclass[a4paper,11pt,preprint]{article}
\usepackage{amsmath}
\usepackage{color}
\usepackage{graphicx}
\usepackage{amssymb}
\usepackage{float}
\usepackage{scrtime}
\usepackage{fancyhdr}
\usepackage{subfigure}
\usepackage{authblk}
\usepackage{lipsum}
\usepackage{rotating}
\usepackage{floatpag}
\usepackage{varioref}
\usepackage{slashed}
\usepackage{enumerate}
\usepackage[toc,page]{appendix}

\usepackage[colorlinks=true
,urlcolor=blue
,anchorcolor=blue
,citecolor=blue
,filecolor=blue
,linkcolor=black
,menucolor=blue
,pagecolor=blue
,linktocpage=true
]{hyperref}

\usepackage[numbers,sort&compress]{natbib}

\setlipsumdefault{131}

\newcommand{\micromegas}{\texttt{MicrOMEGAs}}

\newcommand{\tanb}{\ensuremath{\text{tan} \beta}}

\newcommand{\be}{\begin{equation}}
\newcommand{\ee}{\end{equation}}
\newcommand{\bea}{\begin{eqnarray}}
\newcommand{\eea}{\end{eqnarray}}
\newcommand{\bei}{\begin{itemize}}
\newcommand{\eei}{\end{itemize}}

\newcommand{\neu}[1]{\ensuremath{\tilde{\chi}_{#1}^0}}

\newcommand{\gsim}{\lower.7ex\hbox{$\;\stackrel{\textstyle>}{\sim}\;$}}
\newcommand{\lsim}{\lower.7ex\hbox{$\;\stackrel{\textstyle<}{\sim}\;$}}

\labelformat{subsection}{Section #1} 
\labelformat{equation}{Eq.~(#1)} 
\labelformat{figure}{Fig.~#1} 
\labelformat{subfigure}{Fig.~\thefigure#1} 
\labelformat{table}{Tab.~#1}

\def\be{\begin{equation}}
\def\ee{\end{equation}}
\def\bea{\begin{eqnarray}}
\def\eea{\end{eqnarray}}

\graphicspath{{./figures/}}

\begin{document}

\floatpagestyle{plain}

\pagenumbering{roman}

\renewcommand{\headrulewidth}{0pt}
\rhead{ OHSTPY-HEP-T-14-005\\ MCTP-14-35 }
\fancyfoot{}

\title{\huge \bf{Dark Matter at the Pseudoscalar Higgs Resonance in the pMSSM and SUSY GUTs}}
\date{}
\author[1]{Archana Anandakrishnan\thanks{anandakrishnan.1@osu.edu}
}
\author[2]{Bibhushan Shakya\thanks{bshakya@umich.edu}}
\author[3]{Kuver Sinha\thanks{kusinha@syr.edu}}
\affil[1]{\em Department of Physics, The Ohio State University\newline 191 W.~Woodruff Ave, Columbus, OH 43210, USA \medskip}
\affil[2]{ \em Michigan Center for Theoretical Physics\newline University of Michigan, Ann Arbor, MI 48109, USA \medskip}

\affil[3]{ \em Department of Physics, Syracuse University\newline Syracuse, NY 13244, USA}


\maketitle

\thispagestyle{fancy}


\begin{abstract}\normalsize\parindent 0pt\parskip 5pt

We study dark matter at the MSSM pseudoscalar Higgs resonance (A-funnel), which is one of the few remaining MSSM thermal dark matter candidates in the $100-1000$ GeV range safe from direct detection constraints. To illustrate the various factors at play, this study is performed in two contrasting set-ups: a bottom-up phenomenological MSSM (pMSSM) approach that allows significant freedom, and the top-down, highly constrained Yukawa unified $SO(10)$ GUT model. In the pMSSM,  for $\mu > 0$, the entire parameter space lies above the coherent neutrino background and mostly within reach of XENON1T and LZ, while blind spots exist at $m_A\,\textgreater\, 800\,$GeV for $\mu < 0$; the strongest constraints come from $A/H \rightarrow \tau \tau$ searches at the LHC. For Yukawa unified models, the confluence of $B_s \rightarrow \mu^+ \mu^-$ constraints, fits to the bottom quark and Higgs masses, and gluino mass bounds from the LHC result in a prediction: realizing the pseudoscalar resonance \textit{requires} gaugino mass non-universality, with a mass ratio that is determined to within a small range.

\end{abstract}

\clearpage
\newpage

\pagenumbering{arabic}

\section{Introduction}

Since the weak scale presumably holds the key to a host of pressing theoretical questions such as the exact nature and naturalness of electro-weak symmetry breaking, the WIMP miracle -- the thermal production of a particle with weak-scale mass and interactions in the early Universe matches the observed dark matter relic abundance -- has for long served as a conduit between particle physics model building and dark matter cosmology, and as a target for colliders and direct detection experiments. Central to dark matter model building is the paradigm of supersymmetry, since it comes with its own strong motivations. The most well-studied model of supersymmetry is the minimal one (MSSM), which looks to be increasingly fine-tuned in view of the non-observation of superpartners at the LHC. Cherished simple parametrizations such as the CMSSM or mSUGRA have either been ruled out or pushed into remote corners, and full phenomenological studies of the MSSM (pMSSM) have been undertaken.

The WIMP miracle is realized in the MSSM with the lightest neutralino as the lightest supersymmetric particle (LSP); however, there now exist significant constraints on such a thermal dark matter candidate. Increasingly stringent direct detection bounds, most recently from XENON-100~\cite{Aprile:2012nq} and LUX~\cite{Akerib:2013tjd} are rapidly cutting into the parameter space of supersymmetry, particularly that of well-tempered neutralinos~\cite{ArkaniHamed:2006mb}, requiring fine-tuning to evade such constraints~\cite{Perelstein:2012qg, Perelstein:2011tg}. A pure Higgsino or Wino LSP with annihilation mainly to gauge boson final states can evade such constraints; however, these candidates are constrained by indirect detection~\cite{Easther:2013nga, Allahverdi:2012wb, Cohen:2013ama, Fan:2013faa}. Coannihilation (with slepton, sfermion, chargino, or heavier neutralino) can also result in the correct relic density, but involve a light state close to the LSP mass with stronger interactions that can be probed experimentally.

The purpose of this paper is to study the prospects for the Higgs pseudoscalar resonance (A-funnel), which provides another candidate for thermal dark matter. This involves the lightest neutralino (the Bino) at approximately half the mass of the pseudoscalar Higgs boson $A$, resulting in an enhancement of the annihilation cross-section through $A$-exchange~\footnote{We note that resonance through Z, h, and H are also possible, but do not study these, since they predict larger direct detection cross sections in general. Dark matter at the Higgs resonance is currently constrained more strongly by Higgs invisible decays than by LUX~\cite{Queiroz:2014pra, Queiroz:2014yna}. XENON1T and LZ will significantly constrain this region. There will be complementary constraints in these cases from the LHC.}. We note that it cannot be mostly Wino, nor Higgsino, since annihilations are too efficient already, and the enhancement due to resonance would be detrimental; on the other hand, the annihilation cross section of a Bino is generally too small, leading to over-closure of the Universe upon freeze-out, but the resonance enhancement facilitates the correct relic density. This is a fine-tuned region since the Bino and $A$ masses are generally not correlated; however, given the increasingly stringent constraints from  direct detection experiments, the pseudoscalar resonance constitutes one of the few remaining thermal dark matter candidates that can remain safe from such constraints~\footnote{We will consider only thermal candidates here. For non-thermal histories, several options remain open~\cite{Dutta:2009uf, Allahverdi:2010rh}.}. Moreover, this is a region that can be attacked from multiple angles: new physics contributions to rare decays such as $B_s \rightarrow \mu^+ \mu^-$ and LHC constraints on $A \rightarrow \tau \tau$ search~\cite{Aad:2014vgg, CMStautau} can constrain $m_A$. The combination of direct searches at the LHC, rare $B$-decay processes, and direct detection may therefore ultimately either validate or rule out this window.

We study the prospects of the pseudoscalar Higgs resonance from two different approaches. The first, in Section~\ref{pMSSM}, is from the bottom-up perspective of the phenomenological MSSM with parameters taken to be independent at the weak scale. This offers the opportunity to study the pseudoscalar resonance in isolation with a handful of parameters by decoupling all physics that does not affect it directly. This serves as a general framework within which more constrained models can be embedded. Such a model is studied in Section~\ref{so10guts} in the framework of a Yukawa unified $SO(10)$ GUT model. $SO(10)$ GUTs are one of the economical choices for a grand unified theory and Yukawa unification is a natural feature of these models. In contrast to the pMSSM, a Yukawa unified GUT affords little freedom in its choice of parameters due to simultaneous constraints from fermion masses, flavor physics, and collider bounds. 

The strikingly different characteristics of these two models illustrate some key aspects of the pseudoscalar resonance region. In both extremes, we look for complementary probes of the region and show that the next generation direct detection experiments will be aided by the LHC searches for supersymmetric particles.

\section{Resonance Region in the pMSSM} \label{pMSSM}

As described in the Introduction, the pseudoscalar resonance is an attractive possibility even in the generic MSSM. A detailed study of MSSM resonances was performed previously in~\cite{Hooper:2013qjx, Han:2013gba}. We give projected reaches for upcoming direct detection experiments, examine particularly low cross sections, and discuss complementarity with collider bounds and rare decays.  

To focus on the relevant parameter space, we take a ``decoupling approach", in the sense that any physics that does not contribute to the resonance or constrains it is made heavy enough to decouple from the analysis. Thus we take scalar masses to be uniformly heavy to block stau/stop co-annihilation effects or $t$-channel scalar exchange. The Wino mass is similarly taken to be much larger than the Bino mass to block co-annihilation effects. All these parameters are set to 5 TeV. 

The remaining free parameters, which we scan over using \micromegas~\cite{Belanger:2010pz}, are 
\begin{itemize}
\item Bino mass $m_1$: This is the mass of the dark matter candidate. We scan values between $100-1000$ GeV. Although lower masses are possible, we focus on this range to avoid contamination with the Z and SM-like Higgs resonances, since our main goal is to study the A-funnel. Note that the decoupling limit $m_A\gg m_Z$ holds in this regime. 

 \item Pseudoscalar mass $m_A$: $M_A\sim 2 M_1$ is necessary to enforce the resonance; in our scans, the numbers are constrained to be within $10\%$ of each other.
\item tan$\beta$\,: Scanned between $2\,-\,50$. 
\item $\mu$\,: Since coupling to the pseudoscalar requires a Bino-Higgsino admixture, requiring the correct relic density fixes the value of $\mu$ (we require $\Omega h^2\in(0.08,\,2.0)$). This occurs for values of $|\mu|\sim 1-10$ TeV (see~\cite{Hooper:2013qjx}), and this is our scan interval. While~\cite{Hooper:2013qjx} only studied positive values of $\mu$, we extend our scans to negative values as well.
\end{itemize}

In addition to the correct relic density, we also require the lightest Higgs mass to fall between $123-128$ GeV. While this has no bearing on the neutralino sector or the relic density in the A-funnel, it is necessary to correctly calculate the spin-independent direct detection cross section. 

\begin{figure}[!htp]
\begin{center}
\includegraphics[width=9cm]{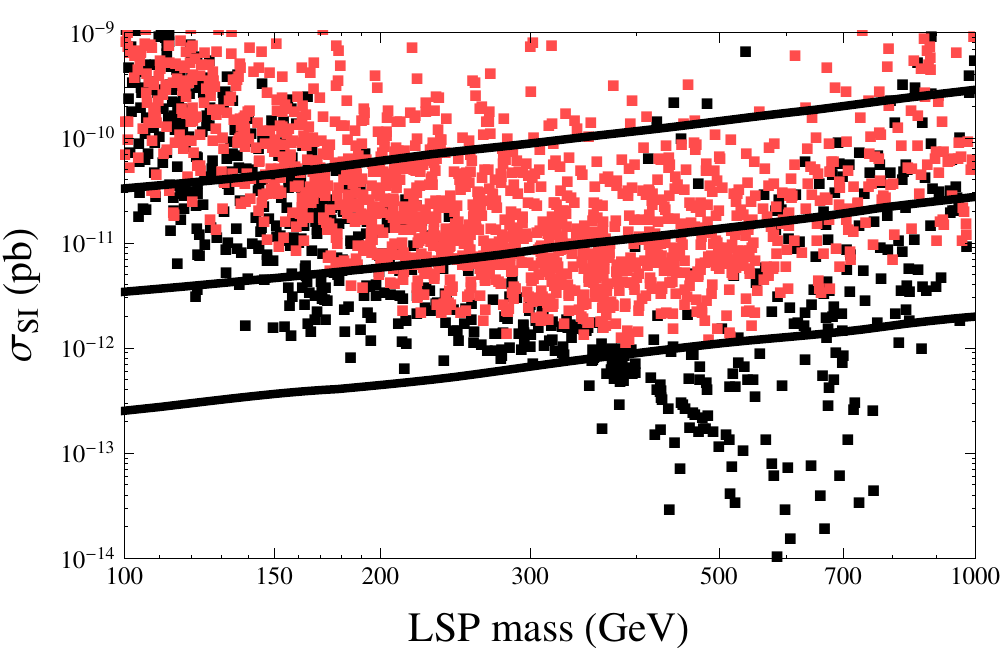}
\caption{\label{mssm1} {\footnotesize {Direct detection cross section as a function of LSP mass. Negative $\mu$ in black, positive in red. The three lines correspond to (top to bottom) the projected XENON1T reach, projected LZ reach, and the irreducible neutrino scattering background.}}}
\end{center} 
\end{figure}

The results of our scan are presented in~\ref{mssm1}, which shows the spin-independent WIMP-nucleon (proton) scattering cross section $\sigma_{SI}$ as a function of the LSP mass. Positive (negative) values of $\mu$ are plotted in red (black). With all scalars heavy, this cross section comes from light and heavy CP-even Higgs exchange in the t-channel, facilitated by the Bino-Higgsino mixture of the LSP that is necessary to obtain the correct relic density. There are also contributions from tree level squark exchange in the s-channel and from gluon loops~\cite{Hisano:2010ct, Cheung:2012qy}, but these are negligible when the superpartners are heavy. The three lines on the plot correspond to (top to bottom) the projected XENON1T reach, projected LZ reach, and the irreducible neutrino scattering background respectively, taken from~\cite{Cushman:2013zza}. 

For positive values of $\mu$ (red points), the cross section is generally found to be around $10^{-11}\,$pb, in agreement with the findings in~\cite{Hooper:2013qjx, Han:2013gba, Kim:2002cy}. Since the resonance requires only a tiny Higgsino fraction in the LSP to give the correct relic density, such small cross sections are a generic feature of this region of parameter space. While the cross sections are well below existing bounds from XENON-100 and LUX, they crucially all lie above the neutrino background, and are therefore within reach of future detectors, although detection will still be challenging.   

In contrast, for negative values of $\mu$ (black points), we find points with cross sections several orders of magnitude below the neutrino background cross section. This was also found in the study in~\cite{Han:2013gba}. As is well-known, such low cross sections occur due to destructive interference between the light and heavy Higgs exchange contributions. Assuming equal couplings to up- and down-type quarks in the nucleus, the cancellation condition can be roughly formulated as~\cite{Feng:2010ef} 
\be
\frac{m_H}{m_h}\sim\sqrt{-\frac{\mu}{m_\chi}\mathrm{tan}\beta} .
\ee
Since this contains precisely the same parameters we have scanned over, this therefore acts as an additional constraint in our parameter space. In the A-funnel set-up, $m_H\sim m_A\sim 2\,m_{\chi}$, so this cancellation condition can be rewritten as
\be
m_A\sim\left(-2\,\mu \,m_h^2\,\mathrm{tan}\beta\right)^{1/3}
\ee\label{cancellationcondition}

For all points with cross sections below $\sim5\times10^{-13}\,$pb, where the cancellation is expected to be in effect, we find that this relation is indeed satisfied within a factor of 2. While a blind spot can in general occur at any LSP mass, its appearance in the case of the A funnel is therefore more strongly constrained.  Requiring the right relic density further constrains a combination of $\mu$ and tan$\beta$: the correct relic density is obtained for $\mu\sim 1-10$ TeV (see e.g.~\cite{Hooper:2013qjx}). Together with Eq.\,\ref{cancellationcondition}, these conditions roughly require $m_A\,\textgreater\,800$ GeV for the blind spot and the pseudoscalar resonance to occur simultaneously; this is indeed visible in \ref{mssm1}.  

\begin{figure}[!htp]
\begin{center}
\includegraphics[width=9cm]{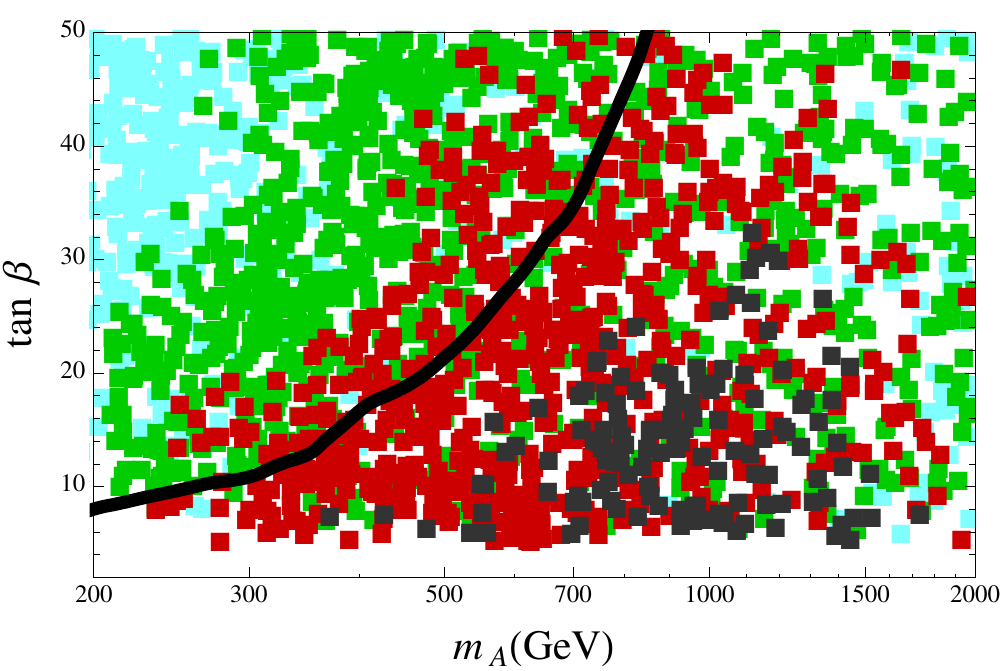}
\caption{\label{mssm3} {\footnotesize {Direct detection cross section color coded in the $m_A-$tan$\beta$ plane. Cyan, green, red and black points correspond to $\sigma_{SI}\in (\textgreater 10^{-10}), (10^{-11},10^{-10}), (10^{-12},10^{-11}), (\textless 10^{-12})$\,pb respectively. The curve denotes the CMS bound from $H/A\rightarrow\tau\tau$~\cite{CMStautau}.}}}
\end{center} 
\end{figure}

Next, we study the scan results in the $M_A-$tan$\beta$ plane. \ref{mssm3} shows the points in this parameter space, color coded by cross section (see caption for details). There exist robust constraints on this parameter space from LHC searches of scalar Higgs boson decays in the $\tau\tau$ channel~\cite{Aad:2014vgg, CMStautau}; the black curve denotes the bound from CMS~\cite{CMStautau}. This rules out a significant chunk of parameter space, in particular pseudoscalar masses below $\sim300$\,GeV. Nevertheless, above this mass, realization of the A-funnel is fairly unconstrained. Future runs of the LHC as well as $e^+e^-$ colliders will further constrain this parameter space (see~\cite{projections} for projections). However, that the cross section tends to fall from left to right shows that points with progressively lower direct detection cross section are also progressively less likely to be ruled out by such searches at the LHC, since both get suppressed at heavier $m_A$. Nevertheless, cross sections sizable enough to be within reach of planned detectors can be realized in all regions of this parameter space, as evident from the presence of blue and green points all over the plot. 

There are additional constraints from $B_s \rightarrow \mu^+ \mu^-$~\cite{Arbey:2012ax} and $B\rightarrow X_s\gamma$~\cite{Carena:2000uj,Ishiwata:2011ab}, but these depend on additional parameters as well as flavor structures and can be evaded by judiciously choosing parameters in the scalar sector (see~\cite{Altmannshofer:2012ks} for extensive discussions and scans). Constraints from metastability of the EW vacuum~\cite{ Kusenko:1996jn, Casas:1995pd} are also important, but can likewise be avoided by properly choosing parameters. These, therefore, do not significantly constrain the A-funnel in the pMSSM due to the freedom in choosing parameter values.

\section{Resonance in the $SO(10)$ GUT Model} \label{so10guts}

In this section, we consider the pseudoscalar resonance in the class of highly predictive Yukawa unified GUT models~\cite{Blazek:2001sb, Baer:2001yy, Blazek:2002ta, Tobe:2003bc, Auto:2003ys, Gogoladze:2011ce, Gogoladze:2011aa, Badziak:2011wm, Anandakrishnan:2012tj, Anandakrishnan:2013cwa, Ajaib:2013zha}. Compared to the pMSSM, we will see that requiring compatibility of the pseudoscalar Higgs resonance with constraints such as $B_s \rightarrow \mu^+ \mu^-$ has non-trivial consequences.

A particularly appealing aspect of $SO(10)$ SUSY GUT models is the possibility of unifying all quarks and leptons of a given generation into a single $16$ representation of the gauge group
\be
W \,\, \supset \,\, \lambda \, 16_3 \, 10 \, 16_3 \,\,.
\ee
Third generation Yukawa unification becomes a possibility for $\tan{\beta} \sim 50$. Yukawa couplings, however, are more sensitive than gauge couplings to weak scale threshold corrections; hence, this unification depends critically on the SUSY spectrum and other parameters. Under the assumption of universal scalar and gaugino masses and $A$-terms at the GUT scale ($m_{16}$, $M_{1/2}$, and $A_0$, respectively)~\footnote{We will explore mild departures from gaugino mass universality in our analysis.}, Yukawa unification prefers a region of parameter space where the following relations hold
\be \label{so10gutrelations}
-A_0 \, \sim \, 2 m_{16}, \,\,\,\,\,\, \mu, m_{1/2} \, \ll \, m_{16} \,\,.
\ee

This region is preferred because large $\tan{\beta}$ corrections to the bottom quark mass are cancelled here. Thus, one can already see that requiring Yukawa unification severely constrains the MSSM parameter space compared to the pMSSM study conducted in the previous section. Further constraints appear as follows:

{\bf (1) $B_s \rightarrow \mu^+ \mu^-$ constraints require $M_A \gsim 1200$ GeV.} When rare $B$ decays are taken into account, there are several immediate consequences. The branching ratio of the decay $B_s \rightarrow \mu^+ \mu^-$ receives large $\tan{\beta}$ enhanced contributions from Higgs-mediated neutral currents, proportional to $A^2_t (\tan^6{\beta})/M^4_A$. The twin requirements of large $\tan{\beta}$ and $A_t$ from \ref{so10gutrelations} necessitate large masses for the Higgs cousins $A^0, H^0,$ and $H^{+}$. Typically, masses larger than $1200$ GeV are preferred in the fits. This puts a lower bound on the neutralino dark matter mass for the resonance to be operational. 

{\bf (2) Heavy scalars with $m_{16} \geq 8$ TeV are required to suppress new physics contributions to $B \rightarrow X_s \gamma$.} Both $B^+ \rightarrow \tau^+ \nu$ and $B \rightarrow X_s \gamma$ receive new physics contributions at large $\tan{\beta}$: the former from charged Higgs bosons through a term that interferes destructively with the SM contribution, the latter from chargino-stop loop and top-charged Higgs loop diagrams. The former is suppressed for decoupled Higgs partners in the same manner as $B_s \rightarrow \mu^+ \mu^-$, and we will not consider it further since $B_s \rightarrow \mu^+ \mu^-$ gives more stringent bounds. For $B \rightarrow X_s \gamma$, the chargino-stop loop term goes as $\sim \frac{\mu A_t}{\tilde{m}^2} \tan{\beta}$, where $\tilde{m}$ is a low-scale squark mass. For large $\tan{\beta}$ and large $A_t$ satisfying~\ref{so10gutrelations}, the universal scalar mass is required to be $\geq 8$ TeV. 

Therefore, a spectrum in the Higgs decoupling limit and with heavy scalars is preferred in Yukawa unified $SO(10)$ when rare $B$ decay constraints are taken into account. Moreover, the spectrum has an inverted mass hierarchy with third family squarks and sleptons between $3 - 6$ TeV. These features, together with the pseudoscalar resonance, have implications for the gaugino sector. To understand these implications, it is instructive to first write down the well known \tanb\ enhanced corrections to the $b$ quark mass 
\be \label{bottom1}
\frac{\delta m_b}{m_b} \simeq \frac{g_3^2}{12 \pi^2} \frac{\mu M_{\tilde{g}} \tanb}{m_{\tilde{b}}^2} + \frac{\lambda_t^2}{32 \pi^2} \frac{\mu A_t \tanb}{m_{\tilde{t}}^2} \,\, .
\ee
To fit data, one needs $\frac{\delta m_b}{m_b}$ to be a negative correction of a few percent. This carries the following implications.

{\bf (3) Light Higgsinos $\lsim 500$ GeV are disfavored.} Given that scalars already have to be heavy, light Higgsinos would further suppress the corrections in~\ref{bottom1} below the nominally required value. In fact, a recent study by some of the authors~\cite{Anandakrishnan:2013tqa} found that for $m_{16} = 20$ TeV, Higgsinos below $\lsim 500$ GeV are disfavored after the recent Higgs mass measurement. Note that negatives values of $\mu$ would be preferable for the expression in \ref{bottom1}, and this choice has been studied in previous works~\cite{Blazek:2001sb}, where it was found that the contributions to $B \rightarrow X_s \gamma$ are enhanced in such scenarios. 
 
{\bf (4) Upper Bound on Gluinos $\sim 2$ TeV}. If gluinos are too heavy, this pushes $A_t$ to larger (more negative) values, beyond maximal mixing, to match the required correction in~\ref{bottom1}. This conflicts with the observed Higgs mass. 

The above observations, taken together, have immediate consequences for thermal dark matter. In~\cite{Anandakrishnan:2013tqa} the well tempered neutralino was explored in this class of models, but found to be increasingly under tension from direct detection. Opting instead for the pseudoscalar resonance, $(4)$ above implies an upper limit on the Bino mass $M_1\sim 300$\,GeV (this follows from the familiar $1:2:6$ mass ratio from gaugino mass unification). Therefore, points $(1)$ and $(4)$ imply {\bf Pseudoscalar Resonance Dark Matter Requires Non-Universal Gaugino Masses.}

We note that this is a non-trivial prediction about the spectrum coming from the requirement of thermal dark matter. Point $(1)$ forces a lower bound $m_{\neu{1}} \gsim 600$ GeV on the dark matter particle, while point $(4)$ makes it clear that the corresponding gluino must be more compressed than the universal case. The requirement of a compressed spectrum was natural in the case of well-tempering, but we find that this requirement holds for the A-resonance region as well. We should note a small caveat to this conclusion. The upper bound on the gluino mass depends on the universal scalar mass $m_{16}$, as is clear from ~\ref{bottom1}. For $m_{16} \, \sim \, 30$ TeV, the upper bound increases to $m_{\tilde{g}} \, \lsim \, 2.8$ TeV. This is still a departure from universality, but less so. For sufficiently high scalar masses and heavy gluino, compatibility with universality may be restored, but such scenarios are not directly testable in the near future.

\subsection{Estimate of Gaugino Mass Ratios}

In this subsection, we carry out a detailed $\chi^2$ analysis to obtain a prediction for the gaugino mass ratio that is preferred by the pseudoscalar resonance.

Non-universality of gaugino masses is parameterized via an additional parameter in the gaugino sector, $\alpha$. The boundary condition we choose for the gaugino masses is mixed modulus-anomalous (mirage) mediation~\cite{Choi:2007ka,Lowen:2008fm,Baer:2006tb}, which is independently well-motivated from string constructions. The gaugino masses at the GUT scale obey a ``mirage" pattern:
\begin{equation} 
M_i = \left(1 + \frac{g_G^2 b_i \alpha}{16 \pi^2} \log \left(\frac{M_{Pl}}{m_{16}} \right) \right) M_{1/2} 
\label{mirage}
\end{equation} 
In the above expression, $\alpha$ controls the relative importance of the universal and anomalous contributions, and $b_i = (33/5, 1, -3) \; {\rm for} \; i = 1, 2, 3$, are the relevant $\beta$-function coefficients. $\alpha = 0$ corresponds to the universal gaugino mass scenario and larger $\alpha$ leads to larger anomaly mediated contributions and a compressed gaugino spectrum. At $\alpha \gtrsim 3$, the gluino becomes the LSP and the spectrum is not viable. Larger $\alpha > 4$ can be accommodated by considering negative $M_{1/2}$ at the GUT scale and yields a Wino LSP~\cite{Anandakrishnan:2013cwa}. 

We calculate 12 low energy observables following the procedure outlined in Ref~\cite{Anandakrishnan:2013cwa}. A global $\chi^2$ analysis is performed with the observables $M_W,\ M_Z,\ G_F,\ \alpha_{em}^{-1},$ $\alpha_s(M_Z),\ M_t,\ m_b(m_b),\ M_\tau, $ $\ b \rightarrow s \gamma,\ BR(B_s \rightarrow \mu^+ \mu^-)$ and $M_{h}$. We then explore the best fit regions and study the thermal relic abundance, $\Omega h^2$.

The input parameters are as follows. There are the three gauge parameters, $\alpha_{G}, M_{G}, \epsilon_3$, where $\alpha_1(M_G) = \alpha_2(M_G) \equiv \alpha_G$, and  $\epsilon_3 = \frac{\alpha_3 - \alpha_G}{\alpha_G}$ is the GUT scale threshold corrections to the gauge couplings. There is one large Yukawa coupling, $\lambda$ which satisfies $\lambda_t(M_G) = \lambda_b(M_G) = \lambda_\tau(M_G) = \lambda$~\footnote{There are typically small corrections to this relation at the GUT scale, coming from the off-diagonal Yukawa couplings to the first two families. In this work, we will mainly consider a third family model, since the details of small off-diagonal Yukawa couplings will not affect the supersymmetric spectrum or dark matter calculations.}. The SUSY parameters defined at the GUT scale are $m_{16}$, $M_{1/2}$, $A_0$, the universal Higgs mass $m_{10}$, and the magnitude of Higgs splitting $D$. There is also the gaugino non-universality parameter $\alpha$. Radiative electroweak symmetry breaking forces non-universal Higgs masses in these models.

We consider several different values of $\alpha$, which will allow us to deviate from gaugino mass universality. For each $\alpha$, we scan different values of $M_{1/2}$ and $\mu$, in the range of $M_{1/2} = 500$ GeV - $950$ GeV, and $\mu = 800$ GeV to $1300$ GeV. We note that the A-funnel can be realized to much higher values of $\mu$, up to several TeV; however, we restrict ourselves to this range since it contains all the qualitatively interesting aspects that we wish to discuss, and higher values of $\mu$ do not introduce any new behavior. The value of $M_A$ is kept within the range of $1200$ GeV - $1400$ GeV.

\begin{figure}[!t]
\begin{center}
\includegraphics[width=8cm]{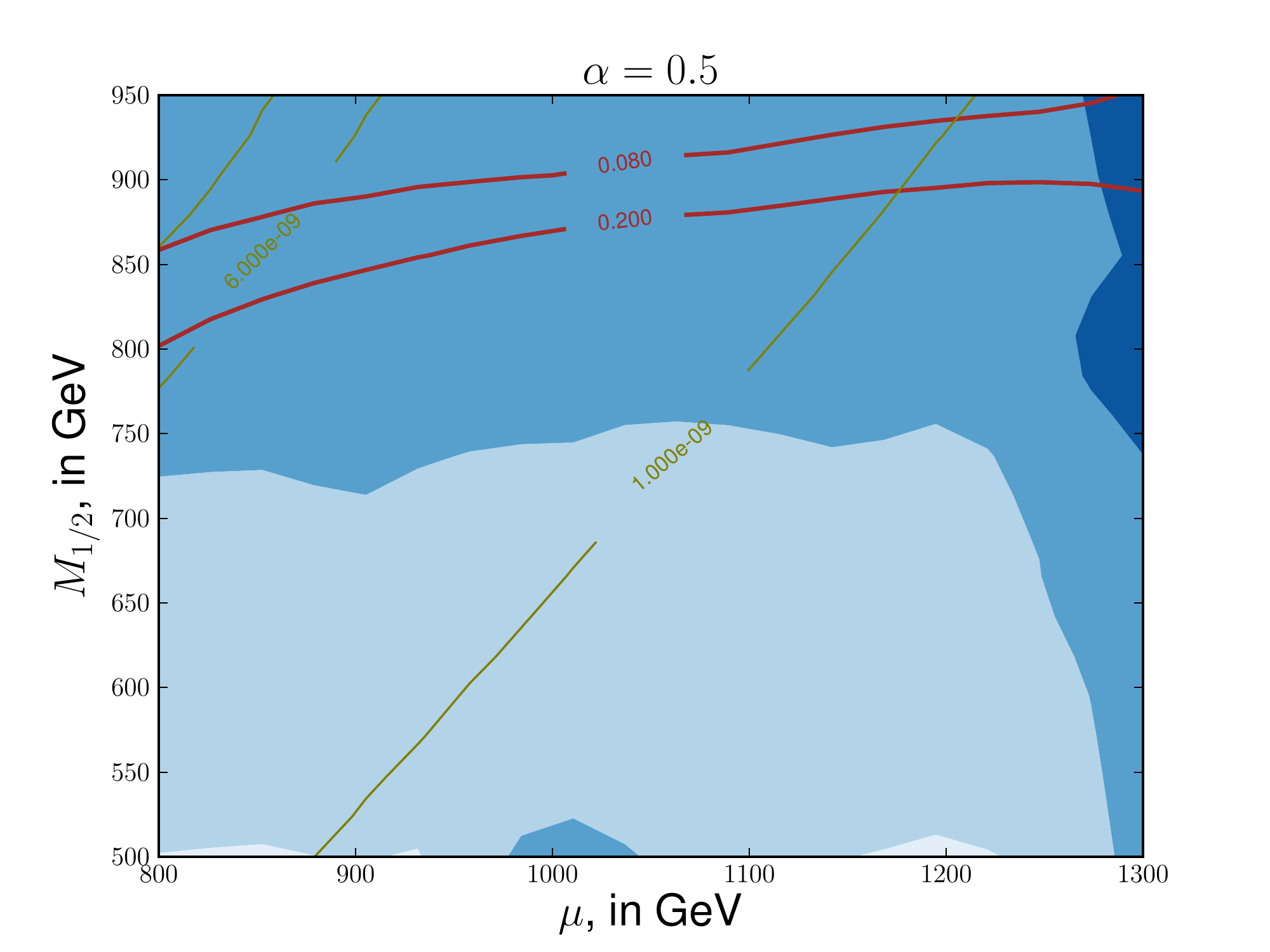}
\includegraphics[width=8cm]{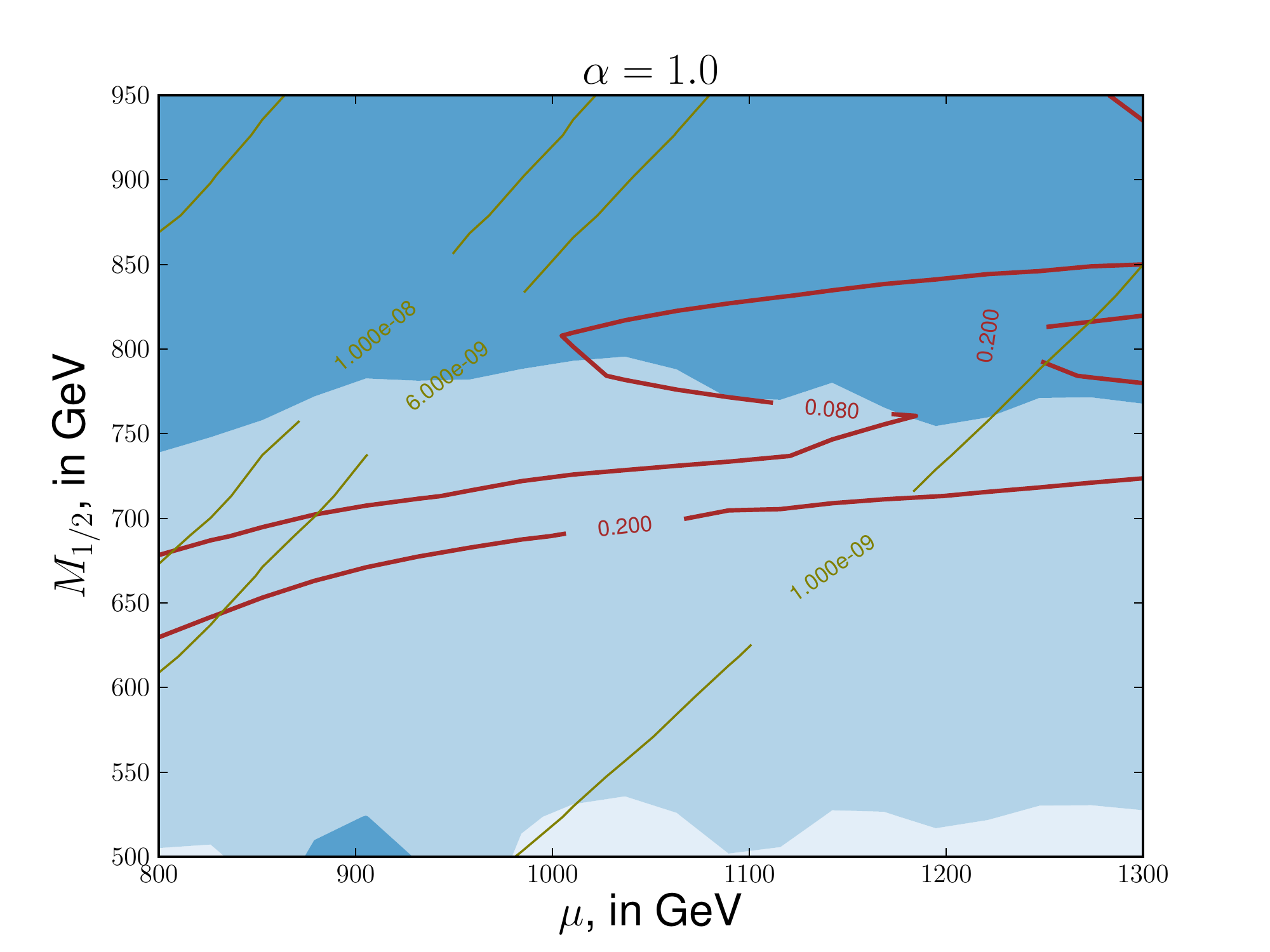}
\includegraphics[width=8cm]{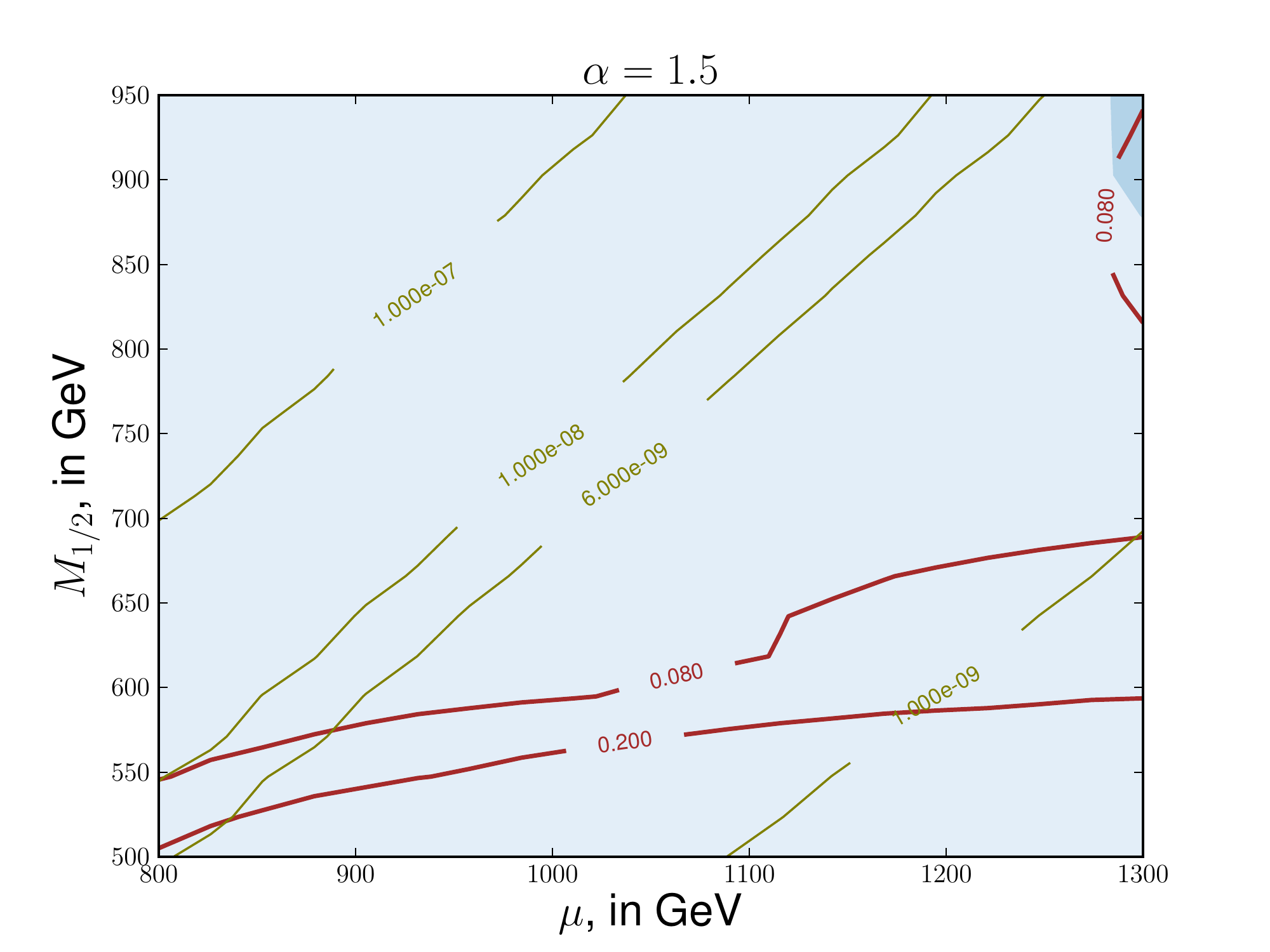}
\includegraphics[width=8cm]{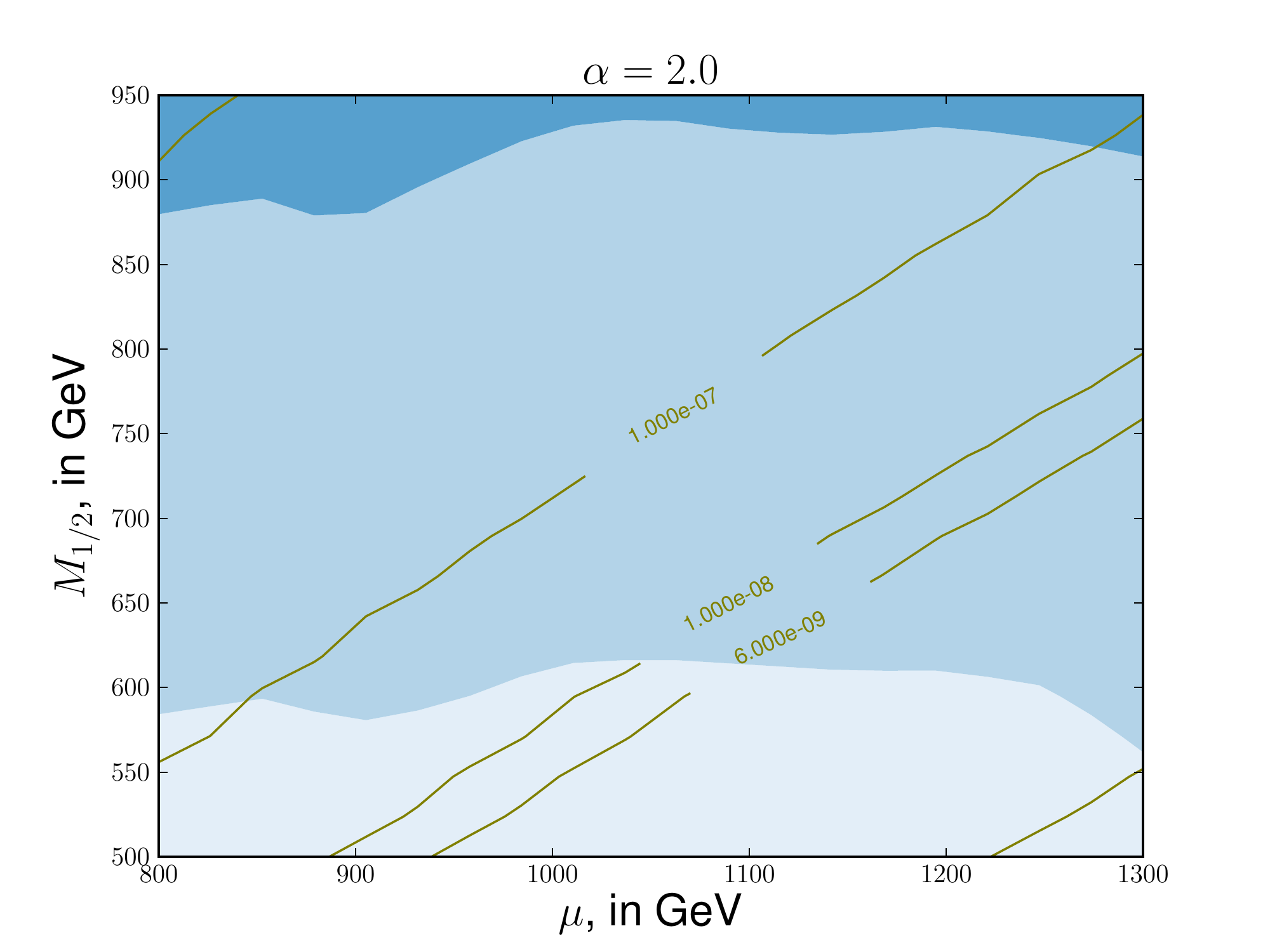}
\caption{\label{afunnelresults} {\footnotesize {\bf Dependence on $\alpha$}: Best fit regions on a graph of $M_{1/2}$ versus $\mu$ in the case of $\alpha = 0.5, 1.0, 1.5$ and $2$. The region between the red lines gives $\Omega h^2 = 0.08 \, - \, 0.2$. The olive contours represent the spin-independent  DM-nucleon direct detection cross-sections. The blue contours (lightest to darkest) represent $\chi^2/dof = 1, 2.3, 3$ and greater.}}
\end{center} 
\end{figure}

We display our results in~\ref{afunnelresults}. The region between the red lines gives $\Omega h^2 = 0.08 \, - \, 0.2$. The olive contours give the spin-independent DM-nucleon scattering cross section.  The scattering cross-section of the DM candidate depends strongly on the Higgsino component. This is evident from~\ref{afunnelresults}, where the scattering cross-sections are the largest where the LSP is a Bino/Higgsino mixture (in the top left corner of the plot) due to the proximity of the values of $M_{1/2}$ and $\mu$. This region shows significantly high DM-nucelon scattering cross section, and is ruled out by current data. The scattering cross section decreases as we go to the bottom right corner of the plot, as Higgsinos progressively become heavier and the Bino/Higgsino mixing decreases. The four shades of blue contours (lightest to darkest) represent $\chi^2/dof < 1, 2.3, 3$, and greater respectively, corresponding to 95\%, 90\%, and 68\% CLs. The darkest shade of blue are the worst fits and we essentially rule the spectrum out as a good solution.

The case of $\alpha = 0$ (universal gaugino masses) is disfavored and hence we do not plot it.
%
%
In the top row of~\ref{afunnelresults}, we display the cases of $\alpha = 0.5$ and $\alpha = 1.0$.  For the case of $\alpha  = 0.5$ which is only a slight deviation from the universal scenario, the relic density corridor lies in the region with $\chi^2/dof > 2.3$ due to a heavy gluino and the Higgs mass is typically less than 120 GeV in this region. The fits become better as $\alpha$ increases, and we see this in the case of $\alpha = 1.0$, where the effect of lower gluino masses pushes the relic density corridor into the 90\% confidence level region.  

In the bottom left panel, we display the case of $\alpha = 1.5$, which is an optimal scenario. For the range of $M_{1/2}$ considered, the gluino mass is in a region that gives good fits to both the bottom mass and the Higgs mass. 
%
Thus, the entire figure has $\chi^2/dof \lsim 1$. The relic density is satisfied in the region which has the pseudoscalar resonance, marked by $M_{1/2} \, \sim \, 500 - 600$ GeV. In the right panel, we display the case of $\alpha = 2.0$. In this case, the anomaly contributions start to dominate, and the LSP becomes predominantly Wino. There is thus no relic density preferred region.

To summarize, we find two effects as we increase $\alpha$. While $\alpha = 0$ does not accommodate good fits to all observables in the A-resonance region, as $\alpha$ increases: $(i)$ the gaugino masses become more compressed and the gluino satisfies condition $(4)$ above; $(ii)$ since the $\beta$-function coefficient of $U(1)_Y$ is large, the resonance is forced to occur at smaller values of $M_{1/2}$. The ideal $\alpha$ region occurs around 1.5, above which the Wino component of the LSP becomes dominant and the gluino mass is also driven below the current LHC bounds. Around this value, we have the following estimate of the gaugino mass ratios at the weak scale:
\be \label{finalratio}
M_{{\rm Bino}} \,\, : \,\, M_{{\rm Wino}} \,\, : \,\, M_{{\rm gluino}} \, \sim \, 1 \, : \, 1.2 \, : \, 2.2 \,\,\,\,\,.
\ee

The robustness of this ratio depends on the value of $\alpha$, which we have demonstrated to be confined within a narrow range for the given selection of $m_A$ and $m_{16}$ on which our results are based. Increasing $m_A$ with $m_{16}$ fixed would require raising the Bino mass, and hence raising $\alpha$ to compress the gaugino spectrum further to accommodate the upper bound on the gluino mass. Requiring the Bino to remain the LSP is found to require that $\alpha$ cannot deviate too much from our optimal value of $\alpha = 1.5$. On the other hand, increasing $m_{16}$ to large enough values may accommodate gaugino universality with $\alpha = 0$, but such scenarios are less testable, as mentioned before. The mass ratios in \ref{finalratio} are therefore reasonably robust within testable models.

In~\ref{benchmarkspectrumbhw}, we display the spectrum and input parameters at a sample benchmark point. The Bino is at $m_{\neu{1}} \sim 666$ GeV, while $M_A = 1300$ GeV, implying that the pseudoscalar resonance is operational here. The rest of the spectrum is very similar to minimal- Yukawa unified GUTs, with heavy scalars. The gaugino spectrum is compressed and the gluino remains the only (but a very strong) viable candidate for detection at the LHC. For direct detection prospects, we refer to \ref{mssm1}. For collider prospects, we refer to~\cite{Anandakrishnan:2013nca, Anandakrishnan:2014nea}. 

In the event of a gluino discovery at the next run of the LHC, \ref{finalratio} can be taken as a hint from the dark sector to aid in our quest for the remaining gaugino spectrum.

\vspace{1cm}
\begin{table}[ht!]
\begin{center}
\scalebox{0.85}{
\renewcommand{\arraystretch}{1.2}
\begin{tabular}{l l r l r l r l r}
\hline
GUT scale & $m_{16}$ & 20408 & $M_{1/2}$ & 650  & $A_0$  & -40656 &
$\alpha$ & 1.5 \\
parameters & $m_{H_d}$ & 27364 & & & $m_{H_u}$ & 24147 & & \\
& 1/$\alpha_G$ & 26.29  & $M_G$ & 2.03 $\times 10^{16}$ & $\epsilon_3$ & 0\%  &
$\lambda$ & 0.585  \\
\hline
 EW parameters & $\mu$ & 1300 & & & tan$\beta$ & 49.13 &  & \\
\hline
Fit & Total $\chi^2$& 1.63 & & & & & & \\
\hline
Spectrum &  $m_{\tilde u}$ & $\sim$ 20103  & $m_{\tilde d}$  & $\sim$ 20203   &
$m_{\tilde e}$  & $\sim$20569 & &  \\
 & $m_{\tilde t_1}$ & 4126  & $m_{\tilde b_1}$  & 5714   & $m_{\tilde \tau_1}$ 
& 8296 & $M_{\tilde g}$ & 1428  \\
& $m_{\tilde\chi^0_1}$   & 666  & $m_{\tilde\chi^0_2}$  &  770  &
$m_{\tilde\chi^0_3}$   & 1304  & $m_{\tilde\chi^0_4}$ & 1309\\
 & $m_{\tilde\chi^+_1}$   & 770  &  &  & $m_{\tilde\chi^+_1}$   & 1309 & & \\
& $M_A$   & 1300  & $M_H^{\pm}$  & 1302 & $M_H$   & 1526 & $M_h$ & 120  \\
 \hline
DM & $\Omega h^2$& 0.151 & & & SI cross-section & & $8.076 \times 10^{-10}$&  \\
\hline 
Dominant $\tilde{g}$ BR & $ tb\widetilde{\chi}^\pm_1$ & 61\% & $t\bar{t}\widetilde{\chi}^0_2$ & 26\% & $ t\bar{t}\widetilde{\chi}^0_1$ & 8 \% & $
b\bar{b}\widetilde{\chi}^0_1 $& 2\% \\
\hline
\end{tabular}}
  \caption{\footnotesize Spectrum at the benchmark point for A-funnel region from the Yukawa unified GUT scenario. The constrained parameter space allows one to determine the entire supersymmetric spectrum. All masses are in GeV and cross-section in pb.}
  \label{benchmarkspectrumbhw} 
\end{center}
\end{table}

\section{Discussion}

Given the freedom of the pMSSM, the A-funnel is readily realized. The spin-independent direct detection cross section is well below current bounds, and the only robust experimental constraint is the heavy neutral Higgs bound from the LHC~\cite{Aad:2014vgg, CMStautau}. There is enough freedom in other sectors of the theory to evade robust bounds from current constraints on $B$-decays. On the direct detection front, a significant chuck of the pseudoscalar resonance region in the pMSSM will be ruled out by the projected reach of XENON1T and especially LZ (see \ref{mssm1}). For $\mu > 0$, the entire parameter space for $100$ GeV $ < \, m_{\neu{1}} \,  < \, 1000$ GeV lies above the coherent neutrino background, so experiments in the future will in principle be able to rule out the resonance. For $\mu < 0$, cancellations can lower the scattering cross section below the coherent neutrino background; however, such cancellations are constrained to a small subset of the parameter space above $m_A\,\textgreater\,800\,$GeV, enforced by additional requirements of the resonance and relic density. 

In the far more constrained Yukawa unified $SO(10)$ GUT models, the story is very different. Unlike the pMSSM, the theory is forced to large values of tan$\beta$, where the bound on $m_A$ from $B_s \rightarrow \mu^+ \mu^-$ becomes extremely strong. Fits to $b-$quark and Higgs masses and gluino mass bounds from the LHC then make it impossible to realize the A-funnel for universal gauging masses, necessitating non-universality. 

The two models therefore paint very different pictures of a thermal dark matter candidate via the pseudoscalar resonance in a supersymmetric model. In the pMSSM, Bino annihilating through the A funnel is a readily available thermal dark matter candidate largely safe from current experimental constraints (and in the case of $\mu\,\textless\,0$, can lie below the neutrino background for direct detection), and largely decoupled from the remainder of the supersymmetric spectrum. In the GUT model, there are extremely strong constraints, but consequently the theory has predictive power. We performed a detailed $\chi^2$ analysis to determine the required degree of non-universality of gauging masses and presented the gaugino mass ratios at the weak scale. The gluino is a very strong candidate for detection at the LHC in this class of models; in the event of such a discovery, these mass ratios could serve as prediction of the Bino and Wino masses if the pseudoscalar resonance is indeed responsible for the observed dark matter density.

\section*{Acknowledgements}


We acknowledge useful conversations with Stuart Raby and Nausheen Shah. We also thank the Mitchell Institute at Texas A\&M, where part of this work was completed, for hospitality. AA is supported by the Ohio State University Presidential Fellowship. AA also thanks the hospitality of CLASSE, Cornell University where a major portion of this work was completed. BS is supported by the DoE under contract DE-SC0007859. KS is supported by NASA Astrophysics Theory Grant NNH12ZDA001N.


\begin{thebibliography}{9}

\bibitem{Aprile:2012nq} 
  E.~Aprile {\it et al.}  [XENON100 Collaboration],
  Phys.\ Rev.\ Lett.\  {\bf 109}, 181301 (2012)
  [arXiv:1207.5988 [astro-ph.CO]].


\bibitem{Akerib:2013tjd} 
  D.~S.~Akerib {\it et al.}  [LUX Collaboration],
  Phys.\ Rev.\ Lett.\  {\bf 112}, 091303 (2014)
  [arXiv:1310.8214 [astro-ph.CO]].


\bibitem{ArkaniHamed:2006mb} 
  N.~Arkani-Hamed, A.~Delgado and G.~F.~Giudice,
  Nucl.\ Phys.\ B {\bf 741}, 108 (2006)
  [hep-ph/0601041].


\bibitem{Perelstein:2012qg} 
  M.~Perelstein and B.~Shakya,
  Phys.\ Rev.\ D {\bf 88}, no. 7, 075003 (2013)
  [arXiv:1208.0833 [hep-ph]].


\bibitem{Perelstein:2011tg} 
  M.~Perelstein and B.~Shakya,
  JHEP {\bf 1110}, 142 (2011)
  [arXiv:1107.5048 [hep-ph]].


\bibitem{Easther:2013nga} 
  R.~Easther, R.~Galvez, O.~Ozsoy and S.~Watson,
  Phys.\ Rev.\ D {\bf 89}, 023522 (2014)
  [arXiv:1307.2453 [hep-ph]].


\bibitem{Allahverdi:2012wb} 
  R.~Allahverdi, B.~Dutta and K.~Sinha,
  Phys.\ Rev.\ D {\bf 86}, 095016 (2012)
  [arXiv:1208.0115 [hep-ph]].


\bibitem{Cohen:2013ama} 
  T.~Cohen, M.~Lisanti, A.~Pierce and T.~R.~Slatyer,
  JCAP {\bf 1310}, 061 (2013)
  [arXiv:1307.4082].


\bibitem{Fan:2013faa} 
  J.~Fan and M.~Reece,
  JHEP {\bf 1310}, 124 (2013)
  [arXiv:1307.4400 [hep-ph]].


\bibitem{Queiroz:2014pra} 
  F.~S.~Queiroz, K.~Sinha and A.~Strumia,
  arXiv:1409.6301 [hep-ph].


\bibitem{Queiroz:2014yna} 
  F.~S.~Queiroz and K.~Sinha,
  Phys.\ Lett.\ B {\bf 735}, 69 (2014)
  [arXiv:1404.1400 [hep-ph]].


\bibitem{Dutta:2009uf} 
  B.~Dutta, L.~Leblond and K.~Sinha,
  Phys.\ Rev.\ D {\bf 80}, 035014 (2009)
  [arXiv:0904.3773 [hep-ph]].


\bibitem{Allahverdi:2010rh} 
  R.~Allahverdi, B.~Dutta and K.~Sinha,
  Phys.\ Rev.\ D {\bf 83}, 083502 (2011)
  [arXiv:1011.1286 [hep-ph]].


\bibitem{Aad:2014vgg} 
  G.~Aad {\it et al.}  [ ATLAS Collaboration],
  arXiv:1409.6064 [hep-ex].
  
  \bibitem{CMStautau}
 [CMS Collaboration],
  CMS-PAS-HIG-13-021.


\bibitem{Hooper:2013qjx} 
  D.~Hooper, C.~Kelso, P.~Sandick and W.~Xue,
  Phys.\ Rev.\ D {\bf 88}, no. 1, 015010 (2013)
  [arXiv:1304.2417 [hep-ph]].


\bibitem{Han:2013gba} 
  T.~Han, Z.~Liu and A.~Natarajan,
  JHEP {\bf 1311}, 008 (2013)
  [arXiv:1303.3040 [hep-ph]].


\bibitem{Belanger:2010pz} 
  G.~Belanger, F.~Boudjema, A.~Pukhov and A.~Semenov,
  Nuovo Cim.\ C {\bf 033N2}, 111 (2010)
  [arXiv:1005.4133 [hep-ph]].


\bibitem{Hisano:2010ct} 
  J.~Hisano, K.~Ishiwata and N.~Nagata,
  Phys.\ Rev.\ D {\bf 82}, 115007 (2010)
  [arXiv:1007.2601 [hep-ph]].


\bibitem{Cheung:2012qy} 
  C.~Cheung, L.~J.~Hall, D.~Pinner and J.~T.~Ruderman,
  JHEP {\bf 1305}, 100 (2013)
  [arXiv:1211.4873 [hep-ph]].


\bibitem{Cushman:2013zza} 
  P.~Cushman, C.~Galbiati, D.~N.~McKinsey, H.~Robertson, T.~M.~P.~Tait, D.~Bauer, A.~Borgland and B.~Cabrera {\it et al.},
  arXiv:1310.8327 [hep-ex].


\bibitem{Kim:2002cy} 
  Y.~G.~Kim, T.~Nihei, L.~Roszkowski and R.~Ruiz de Austri,
  JHEP {\bf 0212}, 034 (2002)
  [hep-ph/0208069].


\bibitem{Feng:2010ef} 
  J.~L.~Feng and D.~Sanford,
  JCAP {\bf 1105}, 018 (2011)
  [arXiv:1009.3934 [hep-ph]].

  \bibitem{projections}
  I.~M.~Lewis,
  arXiv:1308.1742 [hep-ph];
  E.~Brownson, N.~Craig, U.~Heintz, G.~Kukartsev, M.~Narain, N.~Parashar and J.~Stupak,
  arXiv:1308.6334 [hep-ex];
    A.~Djouadi and J.~Quevillon,
  JHEP {\bf 1310}, 028 (2013)
  [arXiv:1304.1787 [hep-ph]];
  S.~Dawson, A.~Gritsan, H.~Logan, J.~Qian, C.~Tully, R.~Van Kooten, A.~Ajaib and A.~Anastassov {\it et al.},
  arXiv:1310.8361 [hep-ex].

\bibitem{Arbey:2012ax} 
  A.~Arbey, M.~Battaglia, F.~Mahmoudi and D.~Martinez Santos,
  Phys.\ Rev.\ D {\bf 87}, 035026 (2013)
  [arXiv:1212.4887 [hep-ph]].


\bibitem{Carena:2000uj} 
  M.~S.~Carena, D.~Garcia, U.~Nierste and C.~E.~M.~Wagner,
  Phys.\ Lett.\ B {\bf 499}, 141 (2001)
  [hep-ph/0010003].


\bibitem{Ishiwata:2011ab} 
  K.~Ishiwata, N.~Nagata and N.~Yokozaki,
  Phys.\ Lett.\ B {\bf 710}, 145 (2012)
  [arXiv:1112.1944 [hep-ph]].


\bibitem{Altmannshofer:2012ks} 
  W.~Altmannshofer, M.~Carena, N.~R.~Shah and F.~Yu,
  JHEP {\bf 1301}, 160 (2013)
  [arXiv:1211.1976 [hep-ph]].


\bibitem{Kusenko:1996jn} 
  A.~Kusenko, P.~Langacker and G.~Segre,
  Phys.\ Rev.\ D {\bf 54}, 5824 (1996)
  [hep-ph/9602414].


\bibitem{Casas:1995pd} 
  J.~A.~Casas, A.~Lleyda and C.~Munoz,
  Nucl.\ Phys.\ B {\bf 471}, 3 (1996)
  [hep-ph/9507294].


\bibitem{Blazek:2001sb} 
  T.~Blazek, R.~Dermisek and S.~Raby,
  Phys.\ Rev.\ Lett.\  {\bf 88}, 111804 (2002)
  [hep-ph/0107097].


\bibitem{Baer:2001yy} 
  H.~Baer and J.~Ferrandis,
  Phys.\ Rev.\ Lett.\  {\bf 87}, 211803 (2001)
  [hep-ph/0106352].


\bibitem{Blazek:2002ta} 
  T.~Blazek, R.~Dermisek and S.~Raby,
  Phys.\ Rev.\ D {\bf 65}, 115004 (2002)
  [hep-ph/0201081].


\bibitem{Tobe:2003bc} 
  K.~Tobe and J.~D.~Wells,
  Nucl.\ Phys.\ B {\bf 663}, 123 (2003)
  [hep-ph/0301015].


\bibitem{Auto:2003ys} 
  D.~Auto, H.~Baer, C.~Balazs, A.~Belyaev, J.~Ferrandis and X.~Tata,
  JHEP {\bf 0306}, 023 (2003)
  [hep-ph/0302155].


\bibitem{Gogoladze:2011ce} 
  I.~Gogoladze, Q.~Shafi and C.~S.~Un,
  Phys.\ Lett.\ B {\bf 704}, 201 (2011)
  [arXiv:1107.1228 [hep-ph]].


\bibitem{Gogoladze:2011aa} 
  I.~Gogoladze, Q.~Shafi and C.~S.~Un,
  JHEP {\bf 1208}, 028 (2012)
  [arXiv:1112.2206 [hep-ph]].


\bibitem{Badziak:2011wm} 
  M.~Badziak, M.~Olechowski and S.~Pokorski,
  JHEP {\bf 1108}, 147 (2011)
  [arXiv:1107.2764 [hep-ph]].


\bibitem{Anandakrishnan:2012tj} 
  A.~Anandakrishnan, S.~Raby and A.~Wingerter,
  Phys.\ Rev.\ D {\bf 87}, no. 5, 055005 (2013)
  [arXiv:1212.0542 [hep-ph]].


\bibitem{Anandakrishnan:2013cwa} 
  A.~Anandakrishnan and S.~Raby,
  Phys.\ Rev.\ Lett.\  {\bf 111}, no. 21, 211801 (2013)
  [arXiv:1303.5125 [hep-ph]].


\bibitem{Ajaib:2013zha} 
  M.~Adeel Ajaib, I.~Gogoladze, Q.~Shafi and C.~S.~Un,
  JHEP {\bf 1307}, 139 (2013)
  [arXiv:1303.6964 [hep-ph]].


\bibitem{Anandakrishnan:2013tqa} 
  A.~Anandakrishnan and K.~Sinha,
  Phys.\ Rev.\ D {\bf 89}, 055015 (2014)
  [arXiv:1310.7579 [hep-ph]].


\bibitem{Choi:2007ka} 
  K.~Choi and H.~P.~Nilles,
  JHEP {\bf 0704}, 006 (2007)
  [hep-ph/0702146 [HEP-PH]].


\bibitem{Lowen:2008fm} 
  V.~Lowen and H.~P.~Nilles,
  Phys.\ Rev.\ D {\bf 77}, 106007 (2008)
  [arXiv:0802.1137 [hep-ph]].


\bibitem{Baer:2006tb} 
  H.~Baer, E.~K.~Park, X.~Tata and T.~T.~Wang,
  Phys.\ Lett.\ B {\bf 641}, 447 (2006)
  [hep-ph/0607085].


\bibitem{Anandakrishnan:2013nca} 
  A.~Anandakrishnan, B.~C.~Bryant, S.~Raby and A.~Wingerter,
  Phys.\ Rev.\ D {\bf 88}, 075002 (2013)
  [arXiv:1307.7723].


\bibitem{Anandakrishnan:2014nea} 
  A.~Anandakrishnan, B.~C.~Bryant and S.~Raby,
  Phys.\ Rev.\ D {\bf 90}, 015030 (2014)
  [arXiv:1404.5628 [hep-ph]].


\end{thebibliography}
\end{document}